# Controllable high-quality electron beam generation by phase slippage effect in layered targets

Q. Yu[1], Y. J. Gu[1,2], X. F. Li[1], S. Huang[1], F. Zhang[1], Q. Kong[1]*, Y. Y. Ma[3], and S. Kawata[4]

[1]*Applied Ion Beam Physics Laboratory, Key Laboratory of the Ministry of Education, Institute of Modern Physics, Fudan University, Shanghai 200433, People's Republic of China*
[2]*Institute of Physics of the ASCR, ELI-Beamlines project, Na Slovance 2, 18221 Prague, Czech Republic*
3College of Science, National University of Defense Technology, Changsha 410073, China
[4]*Department of Advanced Interdisciplinary Sciences, Utsunomiya University, Yohtoh 7-1-2, Utsunomiya 321-8585, Japan*

## Abstract

The bubble structure generated by laser-plasma interactions changes in size depending on the local plasma density. The self-injection electrons' position with respect to wakefield can be controlled by tailoring the longitudinal plasma density. A regime to enhance the energy of the wakefield accelerated electrons and improve the beam quality is proposed and achieved using layered plasmas with increasing densities. Both the wakefield size and the electron bunch duration are significantly contracted in this regime. The electrons remain in the strong acceleration phase of the wakefield while their energy spread decreases because of their tight spatial distribution. An electron beam of 0.5GeV with less than 1% energy spread is obtained through 2.5D PIC simulations.

*Corresponding Author: qkong@fudan.edu.cn

Among the charged particle accelerators that use collective electric fields in plasmas, the laser wakefield accelerator (LWFA) is one of the most promising ideas for high-performance compact electron accelerators [1] because of its higher acceleration gradients ( $E_\parallel(V/cm) \sim \sqrt{n_0(cm^{-3})}$ ) relative to conventional radio frequency accelerating structures. The concept of an LWFA has been investigated for more than three decades. In recent years, with effective injection methods including control of transverse wavebreaking using two counter-propagating lasers [2], magnetic field-assisted self-injection [3], two-color ionization-induced injection [4], colliding pulse injection [5, 6] and density transition injection [7-9], breakthroughs in the generation of quasi-monoenergetic (low energy spread) short bunches of relativistic electrons with energies from MeV to GeV [10, 11] have been achieved in the so-called "bubble regime". However, the electrons' energy spread (a few percent) needs to be further improved for these electrons to see practical application.

In the standard LWFA, the extremely large acceleration gradients of a trailing plasma wave excited by a short laser pulse can intensively accelerate electrons [12]. However, the laser-plasma interaction length limited by the pulse diffraction, the dephasing between the accelerated electron bunch and the plasma wave, and the energy depletion of the laser beam needs to be extended greatly. The diffraction can be overcome by a combination of plasma channel guiding, relativistic self-focusing, and ponderomotive self-channeling [13]. The physics of laser beam propagation in plasmas has been studied in detail [14, 15], and there is ample experimental confirmation of extended guided propagation in plasmas and plasma channels [16, 17]. With the diffraction overcome, another important factor limiting the laser-plasma acceleration length is electron dephasing with respect to the wakefield. To overcome this dephasing problem, tapered plasma waveguides, where the plasma density increases along the longitudinal direction, have been proposed and investigated theoretically in the linear regime [18-20]. Phase synchrony is theoretically realized even in the one-dimensional (1D) configuration with an optimal density profile [19, 20]. However, these schemes are difficult to achieve experimentally because it is difficult to obtain the complicated density profiles [19, 20]. We here propose a new regime extending the electron dephasing length by leveraging the electron phase slippage effect brought on by bubble contraction. Bubble contraction also leads to reduced electron beam duration, so the electron energy spread can be controlled. Layered targets, which can be obtained easily in experiment, can be used as an alternative method to realize this regime.

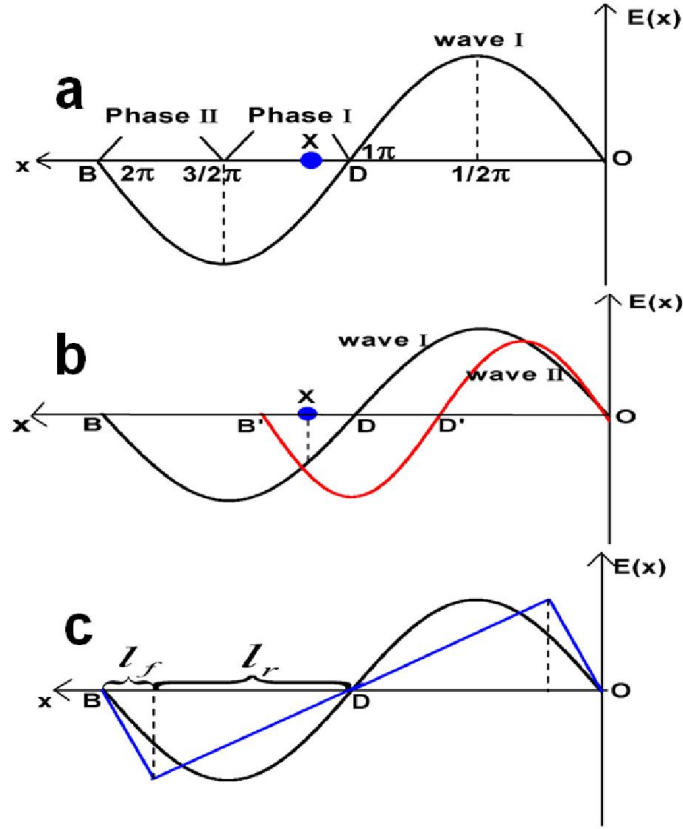

FIG. 1: Schematic diagram of electron phase slippage effect. A normalized wakefield from the bubble regime is presented in (a) and it propagates from left to right. The points O and B denote the front and back of the bubble. The blue point X stands for an accelerated electron. In (b) the wakefield profiles before (black line) and after (red line) contraction are compared. The blue point X also denotes an accelerated electron. The simple harmonic wave (black line) and sawtooth wave are contrasted in (c).

A schematic of a wakefield induced by a short laser pulse propagating in a plasma is shown in Fig. 1 (a). The velocity of the wakefield is the same as the pulse group velocity, $v_g$, and its longitudinal width is the plasma wavelength, $\lambda_p$. Since the accelerated electrons have velocities close to the speed of light, which is greater than $v_g$, the electrons will experience a forward phase shift with respect to the wakefield and reach the dephasing point (marked in Fig. 1 (a) as point D). After reaching the dephasing point, the electrons enter the decreasing phase. However, in an inhomogeneous plasma, the wakefield wavelength will contract when it propagates from a lower-density region to a higher-density region. In this case, the trapped LWFA electrons will remain in the acceleration phase for a longer time and distance. Points O and B in Fig. 1 (a) are the two terminal points of the bubble and point D represents the dephasing point. Here the wakefield propagates from left to right in the $x$-direction and the rest reference frame of the wakefield is employed, which means the right endpoint O remains at rest. We define the phase of a point with $x$-distance from the bubble front as $\Phi_1 = 2\pi x / \lambda_p$. The corresponding phases of the crest and trough are $1/2\pi$ and $3/2\pi$, respectively. The dephasing point has a phase of $\pi$. The electrons located within $\pi \leq \Phi_1 \leq 3/2\pi$, which is marked as Phase I, almost reach the dephasing length. The $3/2\pi \leq \Phi_1 \leq 2\pi$ section is marked as Phase II, which is far from the dephasing point. The red curve in Fig. 1 (b) represents the wakefield profile after contraction, which is called wave II. The phase of the electron at $x$ in wave II can be expressed as $\Phi_2 = 2\pi x / \lambda_{p_2}$, where $\lambda_{p_2}$ is the wavelength of the contracted wakefield. The contraction ratio, which is defined as $\alpha = \Phi_1 / \Phi_2 = \lambda_{p_2} / \lambda_{p_1}$, is less than 1. From the schematic diagram, we can see the same point $x$ experiences a backward phase slippage since the wakefield contracts in the longitudinal direction and the corresponding phase displacement is $\Delta\Phi = 2\pi x (1/\lambda_{p_2} - 1/\lambda_{p_1}) = (1-\alpha) 2\pi x / \alpha\lambda_{p_1}$. This phase slippage makes the electrons at $x$ to be far away from the new dephasing point $D'$ in wave II and extends the dephasing length.

Use of a layered plasma with increasing densities is an alternative method to achieve the phase slippage effect brought about by wakefield contraction. In this kind of target, the plasma wavelength decreases when the bubble propagates from the low to high-density layer, which causes drastic electron phase slippage backward after density transitions.

The trapped LWFA electrons are assumed to be located in Phase I after a sufficient acceleration distance in the first layer. Meanwhile, after the bubble contracts we desire the accelerated electrons to phase slip backward into Phase II. These two points yield:

$$\begin{cases} 1\pi \leq \Phi_1 \leq 3/2\pi \\ 3/2\pi \leq \Phi_2 \leq 2\pi \end{cases} \quad (1)$$

By defining $\Phi = 2\pi x/\lambda_p$ and $\alpha = \Phi_1/\Phi_2 = \lambda_{p_2}/\lambda_{p_1} = \sqrt{n_1/n_2}$, where $n_1$ and $n_2$ are the densities of the low and high-density layers, respectively, in the layered target, Eq. (1) can be expressed as

$$\begin{cases} \pi \leq 2\pi x/\lambda_{p1} \leq 3\pi/2 \\ 3\pi/2 < 2\pi x/\alpha\lambda_{p1} < 2\pi \end{cases} \quad (2)$$

To satisfy expression (2), we have a series of inequalities according to different density scales of plasma layers $\alpha$:

$$\begin{cases} 1/2 \leq x/\lambda_{p1} \leq \alpha & (if \quad \alpha < 2/3) \\ 3\alpha/4 \leq x/\lambda_{p1} \leq \alpha & (if \quad 2/3 \leq \alpha \leq 3/4) \\ 3\alpha/4 \leq x/\lambda_{p1} \leq 3/4 & (if \quad \alpha > 3/4) \end{cases} \quad (3)$$

where $x$ depends on the thickness of the low-density plasma layer $L_1$. In the linear regime, $(c-v_g)L_1/v_g = \lambda_{p_1} - x$. The contraction ratio $\alpha$ is related to the density gap between the two layers of the target by $\alpha = \Phi_1/\Phi_2 = \lambda_{p_2}/\lambda_{p_1} = \sqrt{n_1/n_2}$. So expression (3) is only related to the thickness of the lower-density layer and density scale of the two layers. We can adjust the first layer's thickness and density scale to control the electron phase location. If the electron has already reached the dephasing point before the density transition, *i.e.* $x = \frac{1}{2}\lambda_{p_1}$, the density gap should satisfy $1/2 \leq \alpha \leq 2/3$ and $2.15 \leq n_2/n_1 \leq 4$ to make the electron phase slip back into Phase II after bubble contraction.

With sufficiently high pulse intensities, the interaction between the laser and plasma becomes highly nonlinear. The wakefield profile changes from a simple harmonic (the black line in Fig. 1 (c)) to a sawtooth wave structure (red line in Fig. 1 (c)). The ratio between the widths of the falling edge and the rising edge of the acceleration field is no longer 1:1 (see Fig. 1(c)). We assume the ratio of $l_f / l_r$ is $\varepsilon$, where $l_f$ is the width of the falling edge of the acceleration electric field and $l_r$ is the rising edge width. From nonlinear bubble theory, it is usual that $\varepsilon \ll 1$, and $\varepsilon$ depends on the ratio between the bubble sheath thickness and the bubble radius [22]. In this case, $\alpha$ and $x$ discussed above should be modified nonlinearly. Therefore Eq. (2) and its solution Eq. (3) should be rewritten as:

$$\begin{cases} \dfrac{1}{2} \leq \dfrac{x}{\lambda_1} \leq \dfrac{2+\varepsilon}{2+2\varepsilon} \\ \dfrac{2+\varepsilon}{2+2\varepsilon} \leq \dfrac{x}{\alpha\lambda_1} \leq 1 \end{cases} \quad (4)$$

$$\begin{cases} \dfrac{1}{2} \leq \dfrac{x}{\lambda_{p1}} \leq \alpha & (if\ \alpha < \dfrac{1+\varepsilon}{2+\varepsilon}) \\ \dfrac{2+\varepsilon}{2+2\varepsilon}\alpha \leq \dfrac{x}{\lambda_{p1}} \leq \alpha & (if\ \dfrac{1+\varepsilon}{2+\varepsilon} \leq \alpha \leq \dfrac{2+\varepsilon}{2+2\varepsilon}) \\ \dfrac{2+\varepsilon}{2+2\varepsilon}\alpha \leq \dfrac{x}{\lambda_{p1}} \leq \dfrac{2+\varepsilon}{2+2\varepsilon} & (if\ \alpha > \dfrac{2+\varepsilon}{2+2\varepsilon}) \end{cases} \quad (5)$$

To test the electron phase slippage effect caused by bubble contraction that results in extended dephasing length and reduced electron beam duration, we investigate the interaction of an ultra-short laser pulse with a double-layer plasma by 2.5-dimensional PIC (particle-in-cell) simulations. The simulations were performed using the electromagnetic relativistic code "ZOHAR" [23] by the "move window" technique. The move window size is $250\lambda \times 160\lambda$ with $5031 \times 1068$ cells in the longitudinal and transverse directions; there are eight super particles per cell. The target locates in $10 < x(\lambda) < 1950$, as shown in Fig. 2 (a). The first layer (region 1) is a lower-density plasma at $10 < x(\lambda) < 860$. In the $15 < x(\lambda) < 860$ region is a density plateau with $n_1 = 0.001n_c$ ($n_c$ is the critical plasma density), and there is a linear density up-ramp with the density increasing from 0 to $n_1$ located in $10 < x(\lambda) < 15$. From $860\lambda$ to

$910\lambda$, the plasma density linearly increases from $0.001n_c$ to $0.002n_c$. In the $910 < x(\lambda) < 1950$ region, there is a second layer (region 2) with a uniform density of $0.002n_c$. The laser pulse is linearly polarized in the y-direction with radius $r_0 = 18\lambda$, pulse duration $\tau = 55\,fs$, and laser intensity $I = 2.5\times 10^{20}\,W/cm^2$. The pulse wavelength is $\lambda = 0.8\,\mu m$.

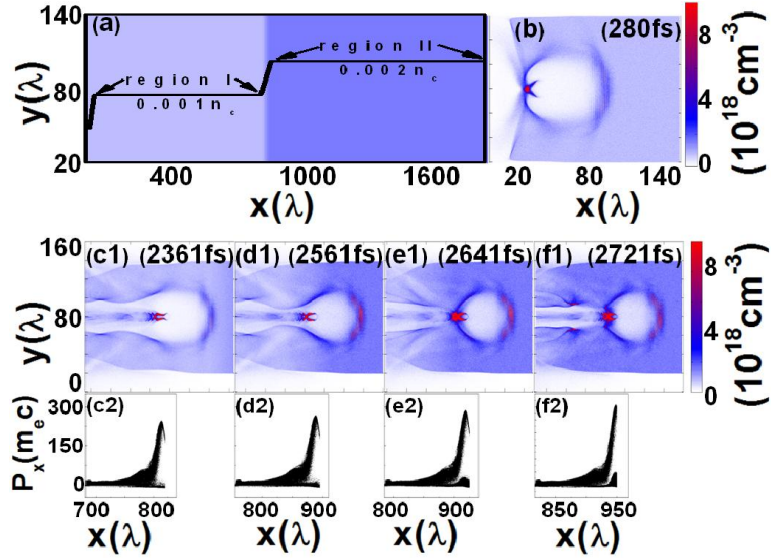

**FIG. 2:** The density outline is presented in (a) for a two-layered target. The electron density profiles are displayed at (b) 280 fs, (c1) 2361 fs, (d1) 2561 fs, (e1) 2641 fs, and (f1) 2721 fs in the two-stage acceleration case. The accelerated electron bunch longitudinal momenta are plotted in (c2), (d2), (e2), and (f2) at 2361 fs, 2561 fs, 2641 fs, and 2721 fs, respectively. The laser parameters are as follows: laser wavelength is 0.8 μm, pulse duration is 55 fs, the radius of the laser pulse is 18λ, and the laser intensity is I=2.125×10$^{20}$W/cm². 

When the following conditions are satisfied, an ultra-intense laser pulse will be self-guided in the plasma. First, the laser power must be above the critical value, $P_c$ [24, 25]

$$P > P_c = 17(n_c/n_e)GW \qquad (6)$$

At the same time, the laser power should be below the upper limit, $P_u$ [26].

$$P < P_u = \frac{n_e r_0^4}{n_c \lambda^4} \times 6.3 TW \qquad (7)$$

where $n_e$ and $n_c$ are the plasma local density and critical density while $\lambda$ and $r_0$ are the laser wavelength and radius, respectively. For a certain laser pulse, the plasma density should exceed the critical value, $n_L$ [26]

$$n_e > n_L = \frac{\lambda^2}{r_0^2} \times 0.044 n_c \qquad (8)$$

The laser employed in our simulations is about 814 TW; however, the upper limit power $P_u$ for $0.001 n_c$ density is 105 TW, which means Eq. (7) is not satisfied in the first layer. The drive pulse is too intense to be focused in a plasma with $0.001 n_c$ density. However, the Rayleigh length for the laser pulse is $z_R = \pi r_0^2 / \lambda \approx 1017 \lambda$, which is much larger than the thickness of the first plasma layer (850λ). Therefore, pulse diffraction effects are not very significant inside the first layer. For the second layer with density $n_e = 0.002 n_c$, we have $P_c = 0.034 GW$, $P_u = 1322 TW$, and $n_L = 0.00016 n_c$, a situation in which Eqs. (6)–(8) are all well satisfied. In this case, the drive pulse is self-focused in the second layer without further diffracting and maintains its intensity at a high value, which is necessary for the realization of long-distance acceleration. A bubble structure is formed at about 280 fs in the first layer and electrons injecting from the bubble's back wall are trapped and accelerated by the wakefield as shown in Fig. 2(b). The first acceleration stage is completed at about 2361 fs when the pulse reaches the end of the first layer. At this time, the accelerated electron bunch has a longitudinal size of about $30 \lambda$ and the high-energy electrons locate in the front of the beam as shown in Figs. 2 (c1) and (c2). One can also find that the bubble shape has transformed to a channel as its rear wall is no longer closed. When the bubble enters the increasing density up-ramp, its back sheath starts to reform to contract its longitudinal size as shown in Fig. 2 (d1). At about 2641 fs, the bubble back sheath is completely closed and an integrated bubble with a smaller longitudinal size is obtained. The accelerated high-energy electrons at the front of the bunch from the first layer are injected into the contracted bubble as illustrated in Figs. 2 (e1) and 2 (e2). These injected electrons continuously phase slip backward until about 2721 fs as displayed in Fig. 2(f1). The reason these electrons continuously phase slip backward at the start of the second plasma layer is discussed below.

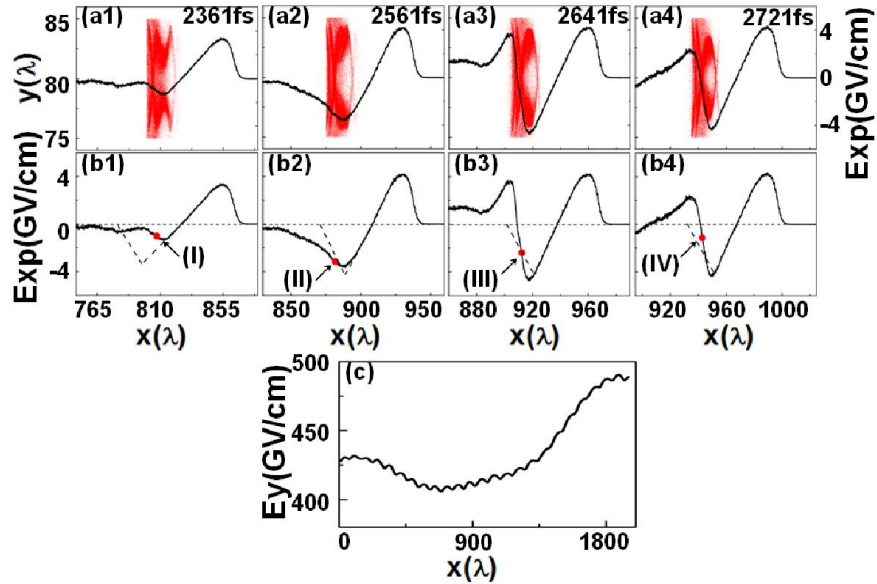

FIG. 3: The longitudinal electric field (E$xp$) along with y=80λ and accelerated electron bunch space distribution at 2361 fs, 2561 fs, 2641 fs, and 2721 fs are plotted in (a1)–(a4), respectively, for the two-layered plasma acceleration case. (b1)–(b4) reveal electron bunch locations relative to the acceleration field at the same times as (a1)–(a4). The red circle points indicate the electrons' average locations at the corresponding time. The peak transverse electric field of the laser pulse as a function of laser location is presented in (c).

The energetic electrons injecting into the contracted bubble will experience backward phase slippage. Figs. 3 (a1) to (a4) show distributions of the accelerated electrons (red dots) and profiles of the longitudinal electric field (black line) along the central axis ($y$=65λ). The strength of the wakefield increases as the bubble propagates from the lower-density plasma to the higher-density plasma. Figs. 3(b1)–3(b4) illustrate the electron beam phase locations (the red circle points I, II, III, and IV) in the acceleration electric fields. The red points are the corresponding electron beam's mean positions in $x$-direction, which represent the electron beam's positions in the wakefield, and the dashed line is the structure of the wakefield obtained by inversion from the positive part of the longitudinal electric field. As demonstrated in Fig. 3 (b1), when the electron beam is close to the end of the first layer at 2361 fs, it locates at $1.33\pi$, which is close to the dephasing point π. If the phase location does not change, the dephasing length will be reached after a short acceleration distance. However, when the bubble enters the density up-ramp, the electron bunch rapidly slips backward with respect to the wakefield, as demonstrated in Fig. 3 (b2), and the corresponding phase position is $1.67\pi$. At 2641 fs, the electron bunch will propagate into the second plasma layer and the electron bunch has slipped to $1.7\pi$ Meanwhile, the intensity of the longitudinal electric field is about four times that in the first layer as shown in Figs. 3 (a3) and (b3).

Furthermore, the electron beam phase slippage does not stop when the bubble enters the second layer. Until about 2721 fs, the phase displacement reaches the maximum and the electron beam locates at $1.75\pi$, as depicted by red circle point IV in Fig. 3(b4). In the linear regime ($a_0 \ll 1$), the normalized bubble phase velocity [27] is

$$\beta_{ph}(x) = \frac{1 - \omega_p^2 / 2\omega_L^2}{1 - (|\xi|/2\omega_p^2)(d\omega_p^2 / dx)} \qquad (9)$$

where $x$ is the propagation direction, $\omega_p$ is the plasma frequency, $\omega_L$ is the laser frequency, and $-\xi = \beta_{gr} ct - x$ denotes the distance from the local position $x$ to the driven laser. When the laser propagates through the up-ramp density gradient, in which $d\omega_p^2 / dx = \omega_L^2 dn / dx > 0$, the normalized bubble phase velocity, $\beta_{ph}$, will exceed unity substantially [27], resulting in electron beam phase slippage backward. When the bubble enters the second layer from the up-ramp gradient, its phase velocity cannot decrease immediately to below unity without some time interval. Further electron bunch phase slippage in the second layer is caused by this non-zero relaxation time. This is also beneficial for further extending the dephasing length and increasing the final electron bunch energy. The evolution of the laser peak transverse electric field as a function of its position is presented in Fig. 3 (c). The plot supports our prediction that the pulse will be self-focused in the second plasma layer. In the first layer, the laser transverse electric field decreases slightly because of pulse diffraction effects. However, the peak amplitude of laser transverse electric field increases from 400 GV/cm to 500 GV/cm inside the second layer by self-focusing.

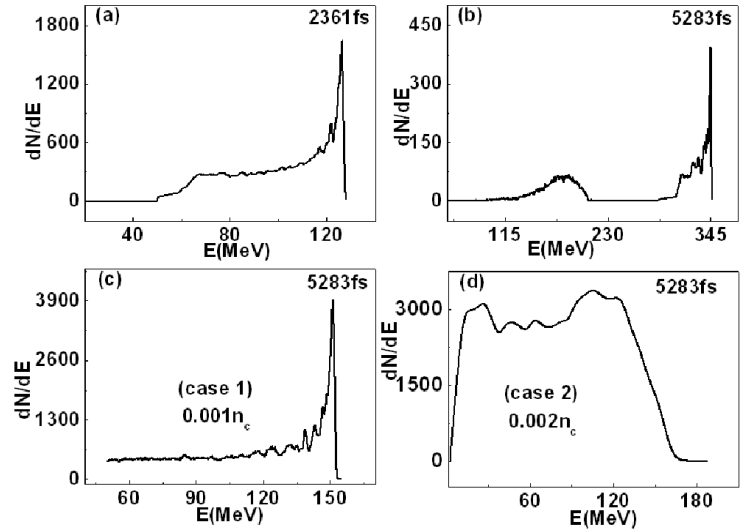

Fig. 4: The accelerated electron bunch energy spectra at the end of the first and second layer at 2361 fs and 5283 fs are shown in (a) and (b) for the two-layered target case. The electron beam energy spectra for the $n_e=0.001n_c$ and $n_e=0.002n_c$ uniform density cases at 5283 fs are also plotted in (c) and (d) with the same laser parameters as that in the two-layered target case.

The accelerated electron bunch energy spectra are plotted in Figs. 4 (a) and (b) after the first and second layer acceleration, respectively. To compare the results with general bubble acceleration in uniform density targets, we also present electron energy spectra for plasmas with uniform density (case 1: $0.001n_c$ and case 2: $0.002n_c$) in Figs. 4 (c) and (d), respectively. From Figs. 4 (b) to (d), one can find in the two-layered target case that the peak electron energy reaches about 340 MeV with effective acceleration gradient of 2.1 GeV/cm, which is 2.27 times and 2.83 times than that in case 1 and case 2, respectively, because of the increase in both dephasing length and wakefield strength. In case 1 and case 2, the corresponding final accelerated electron peak energies are about 150 MeV and 120 MeV, *i.e.*, the effective acceleration gradients are only 0.93 GeV/cm and 0.75 GeV/cm, respectively.

In addition, in the double-layered target acceleration case, the energy spread of the finally accelerated electron bunch is about 0.6%. This monoenergetic beam quality has great potential for application in free electron laser generation. However, in case 1 and 2, the energy spreads are 2% and almost 100%, respectively. There is a marked difference because the accelerated electron bunch is shortened by bubble contraction during the density transition in the two-layered target acceleration case. Meanwhile, the electron charge reduces slightly from $85\,pC/um$ to $76\,pC/um$ after density transition. The duration of the accelerated electron bunch can be controlled by adjusting the bubble contraction ratio depending on the density leap and the up-ramp length between the two uniform density regions, and therefore the corresponding electron bunch will have a low energy spread with appropriate parameters. This is another great advantage of the electron phase slippage effect when compared with the general bubble acceleration scheme.

In the previous layered target case, we have the parameters as $\alpha = \sqrt{n_1/n_2} = 0.71$ and $\varepsilon = l_f/l_r = 0.3$ (read from the wakefield structure). Eq. 5 is therefore simplified as $0.6 < x/\lambda_{p1} < 0.71$. $x/\lambda_{p1}$ is about 0.67 in the simulation, which agrees very well with the theoretical model and supports the model's validity.

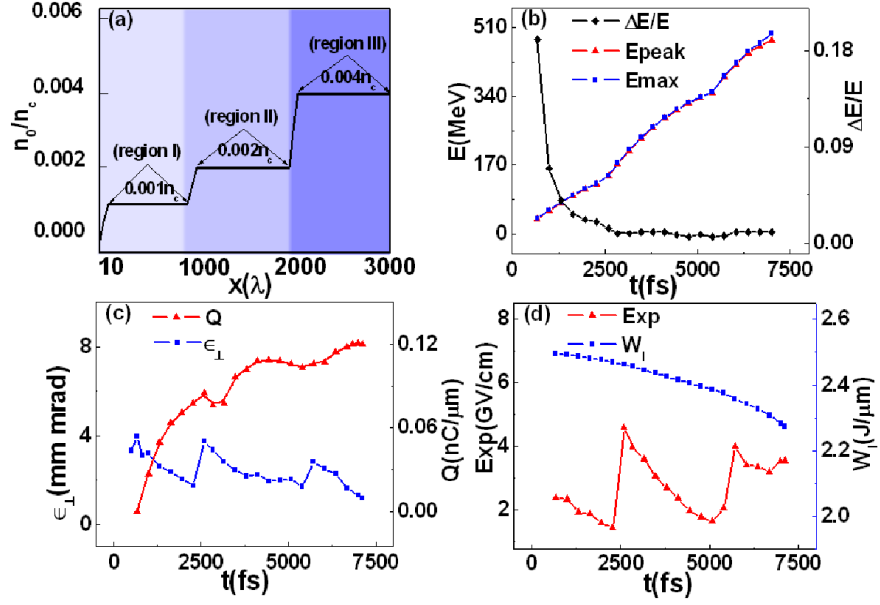

FIG. 5: The density profile of the three-layered target is plotted in (a). In the 10<x(λ)<15 region, the plasma density linearly increases from 0 to 0.001$n_c$. For the 15<x(λ)<860, 910<x(λ)<1950, and 2000<x(λ)<3000 regions, the plasma densities are 0.001$n_c$, 0.002$n_c$, and 0.004$n_c$, respectively. Among them there are two linear up-ramp density gradients located at 860<x(λ)<910 and 1950<x(λ)<2000. The time evolutions of the laser energy and accelerated electron bunch characteristics in the three-layered target acceleration case are shown in (b)–(d). The black diamond, red triangle, and blue square lines in (b) denote the accelerated electron bunch energy spread $\Delta E/E$, peak energy $E_{peak}$, and maximum energy $E_{max}$, respectively. The monoenergetic electron bunch charge Q and transverse emittance $\epsilon_\perp$ are marked by the red triangle and blue square lines in (c). The peak longitudinal electric field $E_{xp}$ and laser energy $W_l$ are displayed as a red triangle and blue square lines in (d). The laser parameters are the same as those in Fig. 2.

When the accelerated electrons reach the dephasing point in the second layer, the phase slippage effect can be achieved once again by adding another plasma layer. We simulated a three-layered target acceleration process with the same laser parameters. The corresponding plasma density profile is shown in Fig. 5 (a). The evolutions of the bunch energy spread, peak, and maximum energies of the accelerated electron beam with time are presented in Fig. 5 (b). The final electron peak energy after experiencing two phase slippages is about 0.5 GeV. According to $\Delta E \sim 48 a_l^{1/2} n_e^{1/2} L$ [21], the theoretical prediction of the final accelerated electron energy should be $\Delta E_1 \sim 48 a_l^{1/2} n_1^{1/2} L_1 = 130 MeV$, $\Delta E_2 \sim 48 a_l^{1/2} n_2^{1/2} L_2 = 210 MeV$, and $\Delta E_3 \sim 48 a_l^{1/2} n_3^{1/2} L_3 = 180 MeV$ for the first, second, and third stage, respectively. The corresponding results in our simulation are $\Delta E_1' =$ 120 MeV, $\Delta E_2' =$ 220 MeV, and $\Delta E_3' = 160$ MeV, which agree with the theoretical estimates very well. We also plot the time evolution of the high-energy electron charge (*E*>50 MeV) and the transverse emittance in Fig. 5 (c). Rather than decreasing, the high-energy electron charge increases slightly in the third acceleration stage instead, finally reaching $120\,pC/um$. This result is because in this period some of the electrons originating from the second and third layers are trapped by the wakefield and join the high-energy electron bunch. The electron beam transverse emittance oscillates around $2.0\,mm\cdot mrad$ during the whole acceleration process. The peak longitudinal acceleration field and the pulse energy as functions of time are plotted in Fig. 5 (d). In each up-ramp gradient stage, the peak acceleration field increases rapidly because of electron accumulation at the back sheath of the bubble. Then it decreases gradually in the plateau density regions because of the electron beam loading effect. The laser energy is depleted slightly after the second stage, and the remaining laser energy and laser power are sufficient to drive and maintain the bubble in the third layer, which benefits from the pulse self-focusing effects from the beginning of the second acceleration period to the end of the whole process. The qualities of the accelerated electron bunch can be improved by controlling the electron phase slippage effect; that is, by controlling the bubble contraction ratio. This can be realized by adjusting the thickness of each density plateau, the density gap between the adjacent plasma layers, and the up-ramp gradient lengths.

In conclusion, we proposed a scheme to extend the electron dephasing length and reduce the electron beam duration by exploiting the electron phase slippage effect. An alternative way to induce electron phase slippage is through plasma wave contraction, which can be realized by a multi-layer target with increasing densities. In this acceleration scheme, the bubble is maintained only by the self-guided laser pulse without any external aids. Because the layered targets have already been obtained in experiments [28], this regime can be implemented easily in experiments. According to our PIC simulation results, the peak energy and energy spread of the accelerated electron beam are both improved by the electron phase slippage effect when compared with the results from general bubble acceleration. After two electron phase slippage events, a 0.5-GeV electron bunch with an energy spread of less than 1% is obtained, though the plasma parameters can be further optimized for better electron quality. The less than 1% energy spread is narrow enough to lend the electrons to applications in many fields including medicine, chemistry, and accelerator physics.

**Acknowledgments**

This work was partly supported by NSFC (No. 11175048), the Shanghai Nature Science Foundation (No. 11ZR1402700), and Shanghai Scientific Research Innovation Key Projects No. 12ZZ011. Support from Shanghai Leading Academic Discipline Project B107, the JSPS-CAS CUP Program, and the CORE of Utsunomiya University is also acknowledged.